\documentclass{Interspeech2024}
\interspeechcameraready

\usepackage{algorithm2e}
\usepackage{cleveref}
\usepackage{float}

\title{On Feature Learning for Titi Monkey Activity Detection}

\name[affiliation={1}]{Aditya}{Ravuri}
\name[affiliation={2}]{Jen}{Muir}
\name[affiliation={1}]{Neil D.}{Lawrence}

\address{
  $^1$University of Cambridge, UK, 
  $^2$Anglia-Ruskin University, UK
}
\email{ar847@cam.ac.uk}

\keywords{voice activity detection, self-supervised learning, representation learning}

\begin{document}

\maketitle

\begin{abstract}
This paper, a technical summary of \cite{ogpaper}, introduces a robust machine learning framework for the detection of vocal activities of Coppery titi monkeys. Utilizing a combination of MFCC features and a bidirectional LSTM-based classifier, we effectively address the challenges posed by the small amount of expert-annotated vocal data available. Our approach significantly reduces false positives and improves the accuracy of call detection in bioacoustic research. Initial results demonstrate an accuracy of 95\% on instance predictions, highlighting the effectiveness of our model in identifying and classifying complex vocal patterns in environmental audio recordings. Moreover, we show how call classification can be done downstream, paving the way for real-world monitoring.
\end{abstract}

\section{Introduction}

Acoustic data analysis provides valuable insights into the ecological, behavioural, and health aspects of animal species. Manual processing of large volumes of acoustic data is challenging, leading to the adoption of machine learning methods in bioacoustic research. This study focuses on Coppery titi monkeys (Plecturocebus cupreus), an accessible species in our local zoos, to explore machine-learning techniques for vocalization analysis. The primary challenge is the development of a framework for voice activity detection using large volumes of passively collected titi monkey data, as relatively small amounts of expert-annotated data is usually available.

In this work, we outline a robust and performant model that appears to be robust and performant in identifying calls, that was first published as part of \cite{ogpaper}. This companion abstract is intended to serve as a technical summary and further exposition on why it was chosen for our use-case.

\section{Activity Detection Methodology}

We break down the problem of activity detection into first modelling the probability of an active call by time segment $\mathbb P(c_t = 1 | \mathbf{a})$ given an audio sequence $\mathbf{a}$, and then finding the $\text{argmax}_{\mathbf{c}} \mathbb P(\mathbf{c} | \mathbf{a})$ to find the most likely activity sequence given the audio.

Initial algorithms that segmented calls using spectrograms with energy between a pre-specified band were successful at identifying calls, albeit with a very high false positive rate, the need to tune hyperparameters based on context (e.g. zoo) and most importantly, led to non-smooth boundaries (e.g. noise can lead to single points in time identified as calls).

\begin{figure}[h]
    \centering
    \includegraphics[width=0.75\columnwidth]{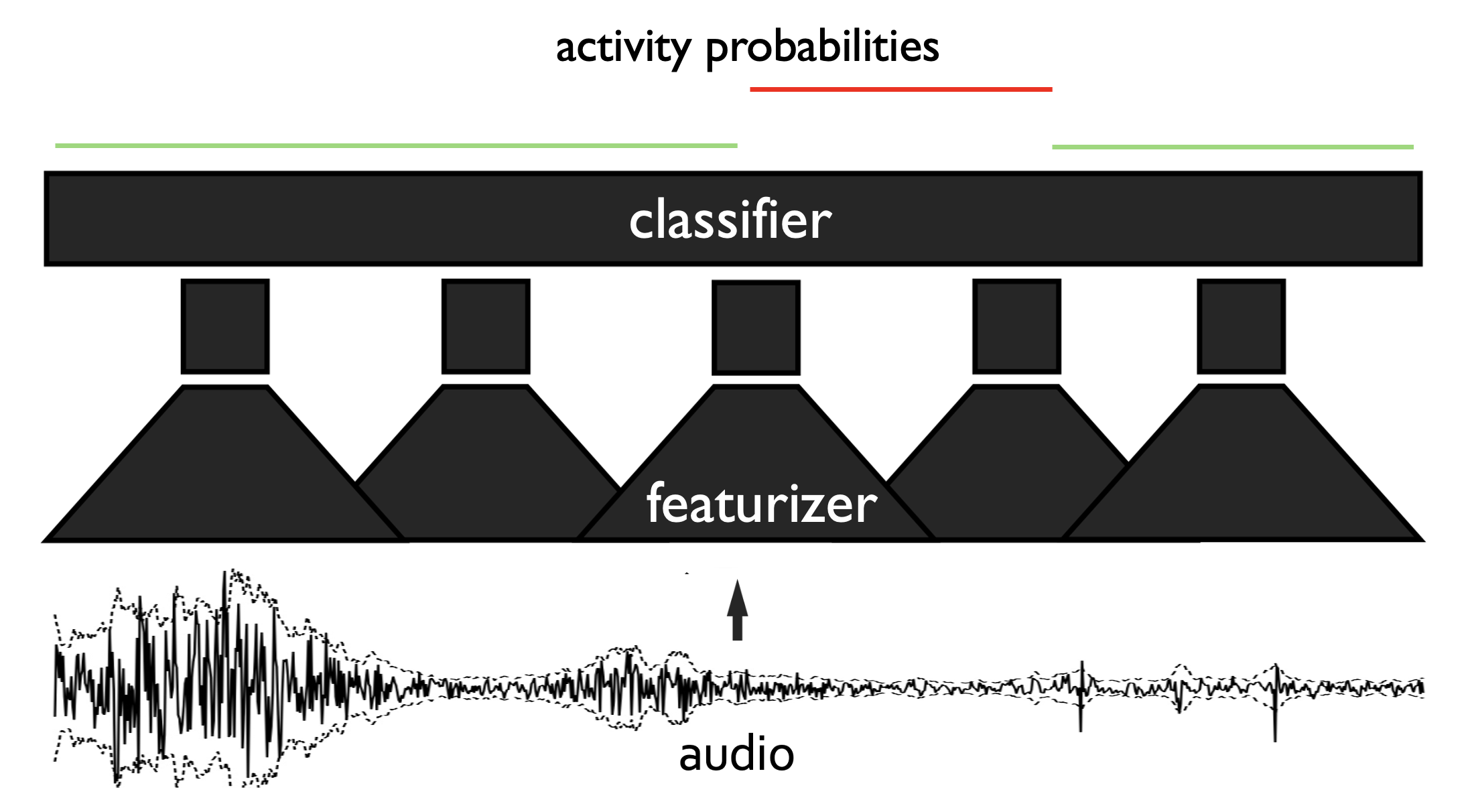}
    \caption{Illustration of model architecture.}
    \label{fig:summary}
\end{figure}

After manually labelling a significant amount of data (but still only a fraction of the collected data), we fit a simplistic model (illustrated in \cref{fig:summary}) to the data, which consisted of around 500 manually labelled files, each of a 10-minute duration.

\textbf{MFCCs are very good representations}

We found that using an MFCC featurizer with 40 MFCCs, used alongside a bidirectional LSTM-based classifier (with three layers and sixteen hidden units, and a single linear layer that compresses the hidden representation to a single probability of activity at an instance in time) works remarkably well at call detection.

The probabilistic model concretely, is, given a segment of five-second $\mathbf{a}$, we model $\forall t: c_t | \mathbf{a} \sim \text{Bernoulli}(\sigma(f(\mathbf{a})_t))$, where $f$ denotes the featurizer and classifier.

We split up our labelled audio into five-second segments, and train the classifier on segments with calls. We achieve around a 95\% accuracy on instance prediction on a validation set (i.e. $\sum_t \mathcal I(c_t = \lceil \sigma(f(a_t)) \rceil) / n \approx 95\% $). The model has a conditional accuracy of about 82\% (i.e. $\sum_{t: c_t = 1} \lceil \sigma(f(a_t)) \rceil/n \approx 82\% $).

\begin{figure}
    \centering
    \includegraphics[width=\columnwidth]{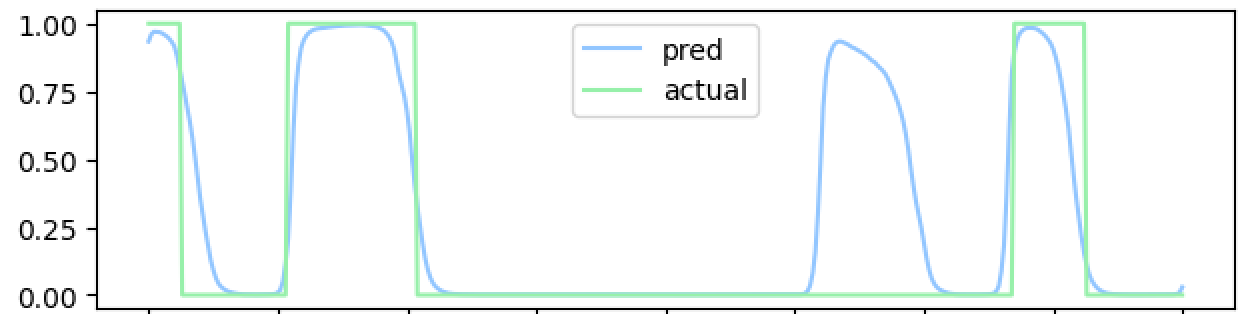}
    \caption{Illustration of model predictions, showing smoothness of output probabilities.}
    \label{fig:smth}
\end{figure}

An interesting consequence of our model architecture is that there's an inductive bias towards smooth outputs, illustrated in \cref{fig:smth}, unlike linear or transformer architectures. We found that using a linear classifier, as is typical with wav2vec (1.0 and 2.0, \cite{wav2vec, wav2vec2}) based ASR models, with either an MFCC or a wav2vec featurizer are non-performant, and a wav2vec featurizer with an LSTM classifier does not perform better than the MFCC version (and is more expensive).

Finally, we use a beam-search decoder implemented in torchaudio \cite{torchaudio21, torchaudio23} to identify segments corresponding to calls (although this is, in practice, similar to a unique consecutive search, it offers the possibility to include a language model as part of the search algorithm).

\textbf{How much data is needed?}

\begin{figure}[H]
    \centering
    \includegraphics[width=\columnwidth]{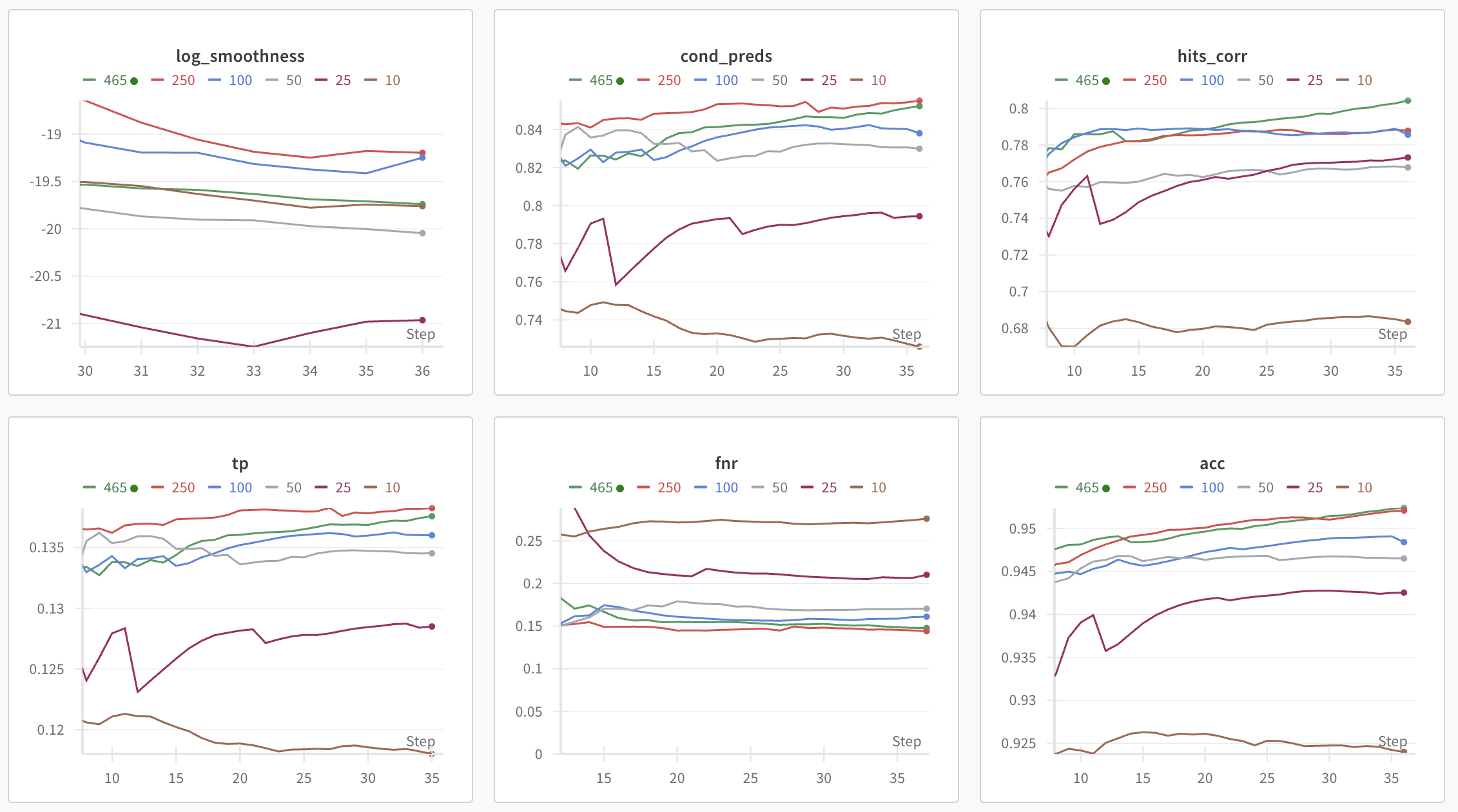}
    \caption{Classification metrics over 1000 epochs of model training, as the number of files used for training are varied.}
    \label{fig:metrics}
\end{figure}

Results of re-training the MFCC-LSTM model on varying amounts of data is shown in \cref{fig:metrics}, showing that model metrics rise significantly until at least 250 files (half of the available data) are used. ``Cond\_preds'' refers to the conditional accuracy of the previous section, and ``hits\_corr'' refers to the correlation between the number of calls identified within one audio segment and the number of calls that were labelled by an expert.

\textbf{If a linear classifier is needed...}

Wav2vec2 based models can be pre-trained (without labels, or using an initial classifier to pick out segments without silence) to obtain classification accuracies of about 92\% (as opposed to 88.95\% using MFCCs), although both of these methods are not very performant.

\section{Call Classification}

\begin{figure}[h]
    \centering
    \includegraphics[width=0.75\columnwidth]{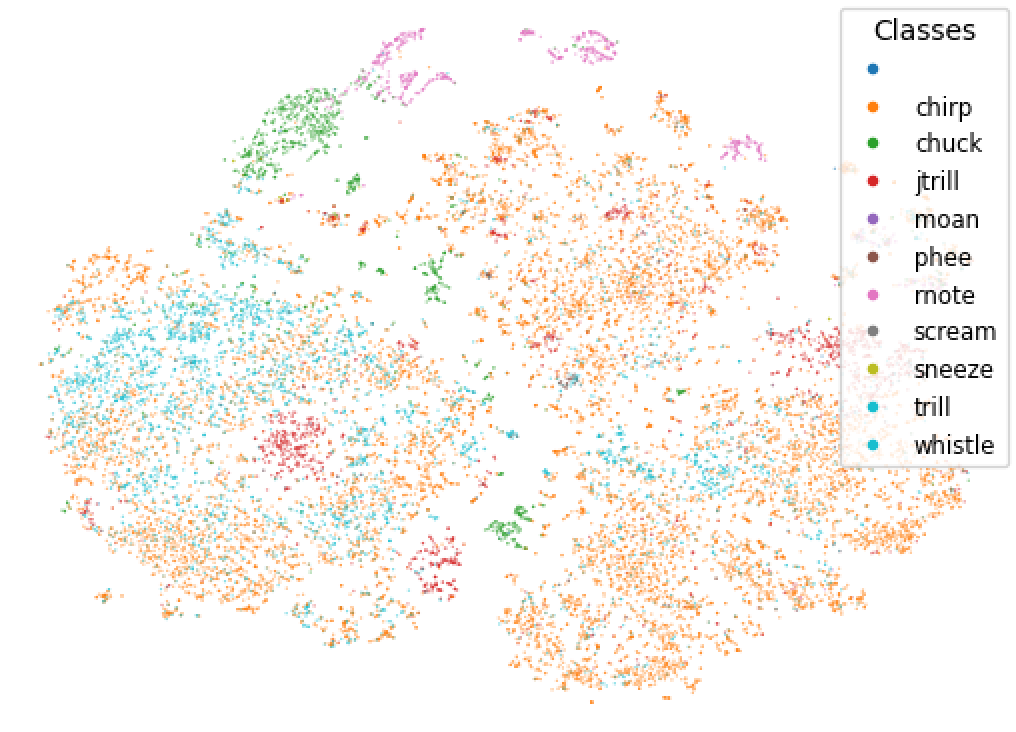}
    \caption{tSNE of average wav2vec-based features coloured by call type. MFCC-based plots look similar.}
    \label{fig:tsne}
\end{figure}

Given identified calls, we found that wav2vec-based features (averaged across the length of the calls) are (somewhat marginally) better than average MFCCs, visualised using a tSNE dimensionality reduction \cite{tsne} in \cref{fig:tsne}.

Moreover, in practice, we're interested in a specific pattern, known as non-linear phenomena (which has been explored for other mammals, for example, \cite{nonlin}). We fit a second classification model to calls identified using our first MFCC-LSTM classifier, to then do a second classification step that grouped identified calls as:
\begin{itemize}
    \item true positive calls without non-linear phenomena
    \item true positive calls \text{with} non-linear phenomena
    \item false positives.
\end{itemize}
The model architecture for the second classifier was similar to our first, although, instead of outputting a label per time point, we do audio level classification, expecting that the second model will only take in calls, and this is done by averaging the latent representation of the LSTM before feeding it into a linear layer. The accuracy of this second step classifier is about 75\%, and when deployed to a totally unseen data file and context, the model hits were all found to be cases of non-linear phenomena, and included cases where the expert found it difficult to identify manually.

\section{Conclusion}

In conclusion, our study presents a highly effective machine learning framework for detecting the vocal activities of Coppery titi monkeys using a combination of MFCC features and a bidirectional LSTM classifier. The model demonstrated a robust capability in reducing false positives and achieving a high accuracy rate in vocal call detection, with promising applications in real-world bioacoustic monitoring. Moreover, the adaptation of our framework to distinguish specific call patterns, including non-linear phenomena, shows potential for enhancing ecological and behavioral studies. This work lays the groundwork for future research in applying advanced machine learning techniques to bioacoustics, potentially extending to a wider range of species and environmental conditions, thereby contributing significantly to wildlife conservation and ecological studies.

\bibliographystyle{IEEEtran}
\bibliography{mybib}

\begin{thebibliography}{1}
\providecommand{\url}[1]{#1}
\csname url@samestyle\endcsname
\providecommand{\newblock}{\relax}
\providecommand{\bibinfo}[2]{#2}
\providecommand{\BIBentrySTDinterwordspacing}{\spaceskip=0pt\relax}
\providecommand{\BIBentryALTinterwordstretchfactor}{4}
\providecommand{\BIBentryALTinterwordspacing}{\spaceskip=\fontdimen2\font plus
\BIBentryALTinterwordstretchfactor\fontdimen3\font minus \fontdimen4\font\relax}
\providecommand{\BIBforeignlanguage}[2]{{%
\expandafter\ifx\csname l@#1\endcsname\relax
\typeout{** WARNING: IEEEtran.bst: No hyphenation pattern has been}%
\typeout{** loaded for the language `#1'. Using the pattern for}%
\typeout{** the default language instead.}%
\else
\language=\csname l@#1\endcsname
\fi
#2}}
\providecommand{\BIBdecl}{\relax}
\BIBdecl

\bibitem{ogpaper}
J.~Muir, A.~Ravuri, E.~Meissner \emph{et~al.}, ``Anonymous paper,'' 2024, under review.

\bibitem{wav2vec}
S.~Schneider, A.~Baevski, R.~Collobert, and M.~Auli, ``{wav2vec: Unsupervised Pre-Training for Speech Recognition},'' in \emph{Proc. Interspeech 2019}, 2019, pp. 3465--3469.

\bibitem{wav2vec2}
A.~Baevski, H.~Zhou, A.~Mohamed, and M.~Auli, ``wav2vec 2.0: A framework for self-supervised learning of speech representations,'' 2020.

\bibitem{torchaudio21}
Y.-Y. Yang, M.~Hira, Z.~Ni, A.~Chourdia, A.~Astafurov, C.~Chen, C.-F. Yeh, C.~Puhrsch, D.~Pollack, D.~Genzel, D.~Greenberg, E.~Z. Yang, J.~Lian, J.~Mahadeokar, J.~Hwang, J.~Chen, P.~Goldsborough, P.~Roy, S.~Narenthiran, S.~Watanabe, S.~Chintala, V.~Quenneville-Bélair, and Y.~Shi, ``Torchaudio: Building blocks for audio and speech processing,'' \emph{arXiv preprint arXiv:2110.15018}, 2021.

\bibitem{torchaudio23}
J.~Hwang, M.~Hira, C.~Chen, X.~Zhang, Z.~Ni, G.~Sun, P.~Ma, R.~Huang, V.~Pratap, Y.~Zhang, A.~Kumar, C.-Y. Yu, C.~Zhu, C.~Liu, J.~Kahn, M.~Ravanelli, P.~Sun, S.~Watanabe, Y.~Shi, Y.~Tao, R.~Scheibler, S.~Cornell, S.~Kim, and S.~Petridis, ``Torchaudio 2.1: Advancing speech recognition, self-supervised learning, and audio processing components for pytorch,'' 2023.

\bibitem{tsne}
\BIBentryALTinterwordspacing
L.~van~der Maaten and G.~Hinton, ``Visualizing data using t-{SNE},'' \emph{Journal of Machine Learning Research}, vol.~9, no.~86, pp. 2579--2605, 2008. [Online]. Available: \url{http://jmlr.org/papers/v9/vandermaaten08a.html}
\BIBentrySTDinterwordspacing

\bibitem{nonlin}
\BIBentryALTinterwordspacing
W.~Fitch, J.~Neubauer, and H.~Herzel, ``Calls out of chaos: the adaptive significance of nonlinear phenomena in mammalian vocal production,'' \emph{Animal Behaviour}, vol.~63, no.~3, pp. 407--418, 2002. [Online]. Available: \url{https://www.sciencedirect.com/science/article/pii/S0003347201919128}
\BIBentrySTDinterwordspacing

\end{thebibliography}

\end{document}